\documentstyle[aps,prl,preprint,floats,epsfig]{revtex}

\begin{document}

\preprint{\tighten\vbox{\hbox{\hfil CLEO CONF 02-05}
                        \hbox{\hfil ICHEP ABS947}
}}

\title{Search for $\eta_{b}(1S)$\\
in\\
Inclusive Radiative Decays of the $\Upsilon (3S)$}

\author{CLEO Collaboration}
\date{July 19, 2002}

\maketitle
\tighten

\begin{abstract} 
We have searched for the bottomonium $\eta_b (1S)$ via the hindered magnetic dipole (M1) photon transition $\Upsilon(3S)\to$ $\gamma \eta_{b}(1S)$.
No evidence for such a transition is found in the data sample of $4.7\times 10^6$ $ \Upsilon (3S)$'s collected with the CLEO-III detector. 
We set upper limits on the branching ratio from 9.30 to 9.43 GeV/$c^2$ of $\eta_{b}(1S)$ masses. These upper limits rule out many previously published phenomenological estimates of the rate for this transition.
\end{abstract}

\vfill
\begin{flushleft}
.\dotfill .
\end{flushleft}
\begin{center}
Submitted to the 31$^{\rm st}$ International Conference on High Energy
Physics,\\ July 2002, Amsterdam
\end{center}

\newpage

{
\renewcommand{\thefootnote}{\fnsymbol{footnote}}


\begin{center}
A.~H.~Mahmood,$^{1}$
S.~E.~Csorna,$^{2}$ I.~Danko,$^{2}$
G.~Bonvicini,$^{3}$ D.~Cinabro,$^{3}$ M.~Dubrovin,$^{3}$
S.~McGee,$^{3}$
A.~Bornheim,$^{4}$ E.~Lipeles,$^{4}$ S.~P.~Pappas,$^{4}$
% ...
\end{center}
 
\small
\begin{center}
$^{1}${University of Texas - Pan American, Edinburg, Texas 78539}\\
$^{2}${Vanderbilt University, Nashville, Tennessee 37235}\\
$^{3}${Wayne State University, Detroit, Michigan 48202}\\
$^{4}${California Institute of Technology, Pasadena, California 91125}\\
% ...
\end{center}

\setcounter{footnote}{0}
}
\newpage

\section{Introduction}
It has been more than 20 years since the $\Upsilon$(1S) was discovered \cite{ups1s}, yet no singlet state of $b\bar{b}$ system has been seen, including the ground state of the $b\bar{b}$ system, $\eta_{b}$(1S).

Singlet S states ($\eta_b$) of bottomonium may be produced in the decays of triplet S states ($\Upsilon$) through magnetic dipole (M1) radiative transitions which occur between states that have opposite quark spin configurations and the same orbital angular momentum. The decay rate for magnetic dipole transitions between between $\Upsilon$ and $\eta_{b}$ is given by \cite{Godfrey}

\begin{equation}
\Gamma[\Upsilon(nS) \to\eta_{b}(n'S)\gamma]=\frac{4}{3} \alpha \frac{e_b^2}{m_b^2}  I^2 k^3
\end{equation}

\noindent where $\alpha$ is the fine-structure constant, $e_{b}$ is the b-quark charge (-$\frac{1}{3}$), $I$ is the model dependent interaction matrix element, $k$ is the photon energy, and $m_{b}$ is the b-quark mass in GeV/$c^2$.

There are two classes of M1 transitions: \emph{direct}, when 
the initial and the final radial quantum numbers are the same ($n=n'$), and \emph{hindered}, when they are not the same ($n\ne n'$).
Small, non-zero values of $I$ for hindered transitions arise from either hyperfine interactions
or finite size corrections, as well as other relativistic effects.
Estimates of these effects are model dependent.

As Godfrey and Rosner  \cite{Godfrey} have pointed out, some hindered transitions could have observable 
rates, in spite of smaller matrix element $I$, because of the large phase space factor, $k^3$. 
In addition, photons emitted in direct transitions are expected to be very soft ({\it i.e.} $k<100$ MeV for $\Upsilon (nS) \to \gamma \eta_b (nS)$) and
fall into the region where experimental backgrounds are high and energy resolution is poor.
This makes detection of direct M1 transitions in the $\Upsilon$ system very difficult.
From the standpoint of $\gamma$ detection and energy resolution, we have a better chance of detecting hindered transitions.

\begin{figure}[!h]
\begin{center}
\scalebox{.80}{\epsfig{file=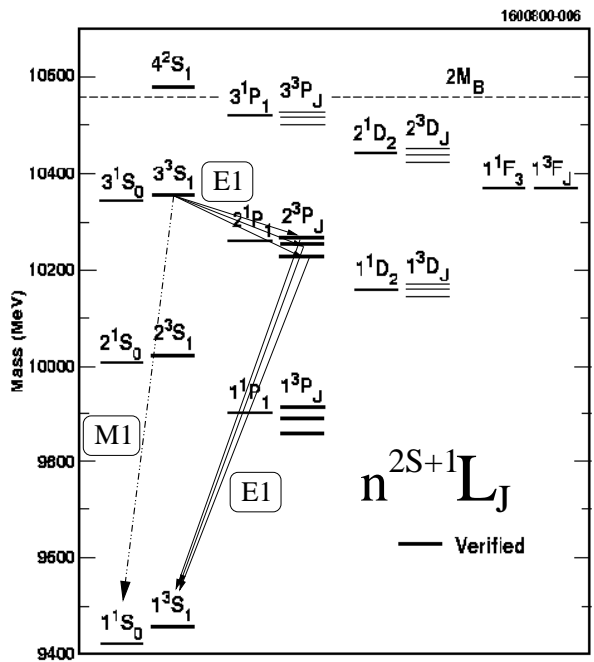,width=4.0in}}
\caption{The $\Upsilon$ energy level spectrum.}
\label{fig:states}
\end{center}
\end{figure}

In this paper we present the result of a search for the hindered magnetic dipole (M1)
transition $\Upsilon$(3S) $\to$ $\gamma \eta_{b}$(1S) using the CLEO III detector.
In Section \ref{sec:event}, we describe the detector, data sample and event selection used for this search. 
Section \ref{sec:e1} is devoted to the electric dipole (E1) transition, 
$\chi_{b}(2P_{J})\to \gamma \Upsilon(1S)$, whose emitted photon energy is 
$\approx$770 MeV. This is close to the predicted photon energy ($\approx$910 MeV \cite{Godfrey}) in the 
$\Upsilon (3S)\to \gamma \eta_{b}(1S)$ transition (see Figure \ref{fig:states}).
We use this nearby E1 transition to optimize our cuts for the signal photons.
In Section \ref{sec:fit}, we describe how we fit to and extract the signal candidate from the data. 
Section \ref{sec:eff} describes the photon detection efficiency and possible
systematic errors in our calculation of branching ratios.
Since no convincing signal is found, 
we present upper limits on the branching ratio.
We summarize our results in section \ref{sec:concl}.

\section{CLEO-III Data Sample, Detector and Event Selection}\label{sec:event}

Recently the CLEO III detector has accumulated an integrated luminosity of 1.1 fb$^{-1}$ using 
the Cornell Electron Storage Ring (CESR) running at the $\Upsilon (3S)$ resonance 
and 0.13 fb$^{-1}$ in the continuum (below the $\Upsilon (3S)$ resonance).
This is roughly a ten-fold increase in statistics over $\Upsilon (3S)$ data collected with the CLEO II detector. While several changes have been made to the detector in the upgrade from CLEO II to CELO III \cite{cleo3}, the CsI calorimeter, upon which this analysis depends, has undergone only slight modification.
The CsI calorimeter barrel section is unchanged with respect to CLEO II, 
and the endcaps have been rebuilt to accommodate the new CESR interaction quadrupole magnets.
Material in front of the calorimeter was significantly reduced for the barrel ends and the endcap sections,
improving photon efficiency and energy resolution.

Since $\eta_b$ is not expected to have a large branching ratio to any particular exclusive
final state, we look inclusively for the photon from $\Upsilon (3S) \to \gamma \eta_b(1S)$.
We first select generic hadronic events by requiring the following criteria; (i) number of charged tracks $>$ 2, (ii) total visible energy\footnote{Total energy of charged tracks and calorimeter energy of showers that do not match with charged tracks} $>0.2\,E_{CM}$, (iii) number of charged tracks $>$ 4  or total CsI calorimeter energy $>0.15\,E_{CM}$ and either CsI calorimeter energy or twice the energy of the most energetic shower in calorimeter $<0.75\,E_{CM}$.                
We then look at the inclusive photon spectrum, searching for a peak in the $\approx900$ MeV range.
 
According to our Monte Carlo in which generic $\Upsilon$(3S) decays were generated by LUND/JETSET,
 our hadronic event selection efficiency is 93.2\%. The inefficiency here is mostly due to the leptonic decays of $\Upsilon$(3S) and small contributions due to the cascade decays of $\Upsilon$(3S) to $\Upsilon$(2S) or $\Upsilon$(1S) which subsequently decay into two leptons.
We estimate the number of $\Upsilon$(3S) produced in CLEO III to be 4.7$\times10^{6}$ by correcting the number of continuum subtracted hadronic events by this efficiency.

\section{Cut optimization on the Electric Dipole (E1) Transition}\label{sec:e1}

To optimize our event selection, we use the nearby structure of the Electric Dipole (E1) transition, 
$\chi_{b}(2P_{J})\to \gamma \Upsilon (1S)$ (see Figure \ref{fig:states}). We optimize our cuts such that these E1 peaks in our data sample are best resolved. In this section, we first describe our selection cuts used to select isolated photons. 
Then we describe our efforts to suppress $\pi^{0}$'s, which are the dominant background in this analysis.

\subsection{Isolated Photon Selection Criteria}

We select photons only in the barrel region ($\vert \cos\theta\vert<0.804$),
\footnote{$\theta$ is the polar angle measured relative to the beam pipe.}
which offers high efficiency and the best energy resolution. We reduce this sample to include only well-isolated photons by requiring that the ratio of the two sums of energies in the group of 9 and 25 crystals around the central crystal should peak close to 1. We set the cut to be the value determined by Monte Carlo such that the cut retains 99\% of the generated photons. We further reject hadronic showers and showers due to electrons by requiring that they do not match charged tracks.

\subsection{$\pi^{0}$ Suppression}
In any analysis of inclusive photon spectra, the background arises primarily from $\pi^{0}$'s. In order to suppress this background, we pair our photon candidates with other neutral showers in either the endcap ($\vert \cos\theta\vert>0.849$) or barrel region\footnote{We exclude the transition region between the endcap and barrel regions} and reject the candidate of the pair that satisfies the following $\pi^{0}$ criteria.
We first cut on $\vert \sigma_{\pi^{0}}\vert=\vert (m_{\gamma\gamma}-m_{\pi^{0}})/\Delta m_{\gamma\gamma}\vert <2.5$ where $m_{\gamma\gamma}$ is an invariant mass of the two photons and $\Delta m_{\gamma\gamma}$ is its estimated resolution. When $\vert \sigma_{\pi^{0}}\vert<2.5$ we require the cosine of the opening angle between the two photons to be greater than 0.7.

\section{Fitting the data}\label{sec:fit}

We fit the inclusive photon spectrum with three components;
(i) a 3rd order exponential polynomial for background, 
(ii) a peak due to $\chi_b(2P_J)\to\gamma\Upsilon(1S)$ represented by three overlapping Gaussians due to different members of the  $\chi_b(2P_J)$ triplet, and
(iii) a signal peak due to $\Upsilon(3S)\to\gamma\eta_b(1S)$ represented by a Crystal Ball Line Shape, which is a Gaussian with an asymmetric low energy tail.

\begin{figure}[ht]
\begin{center}
\scalebox{0.8}{\epsfig{file=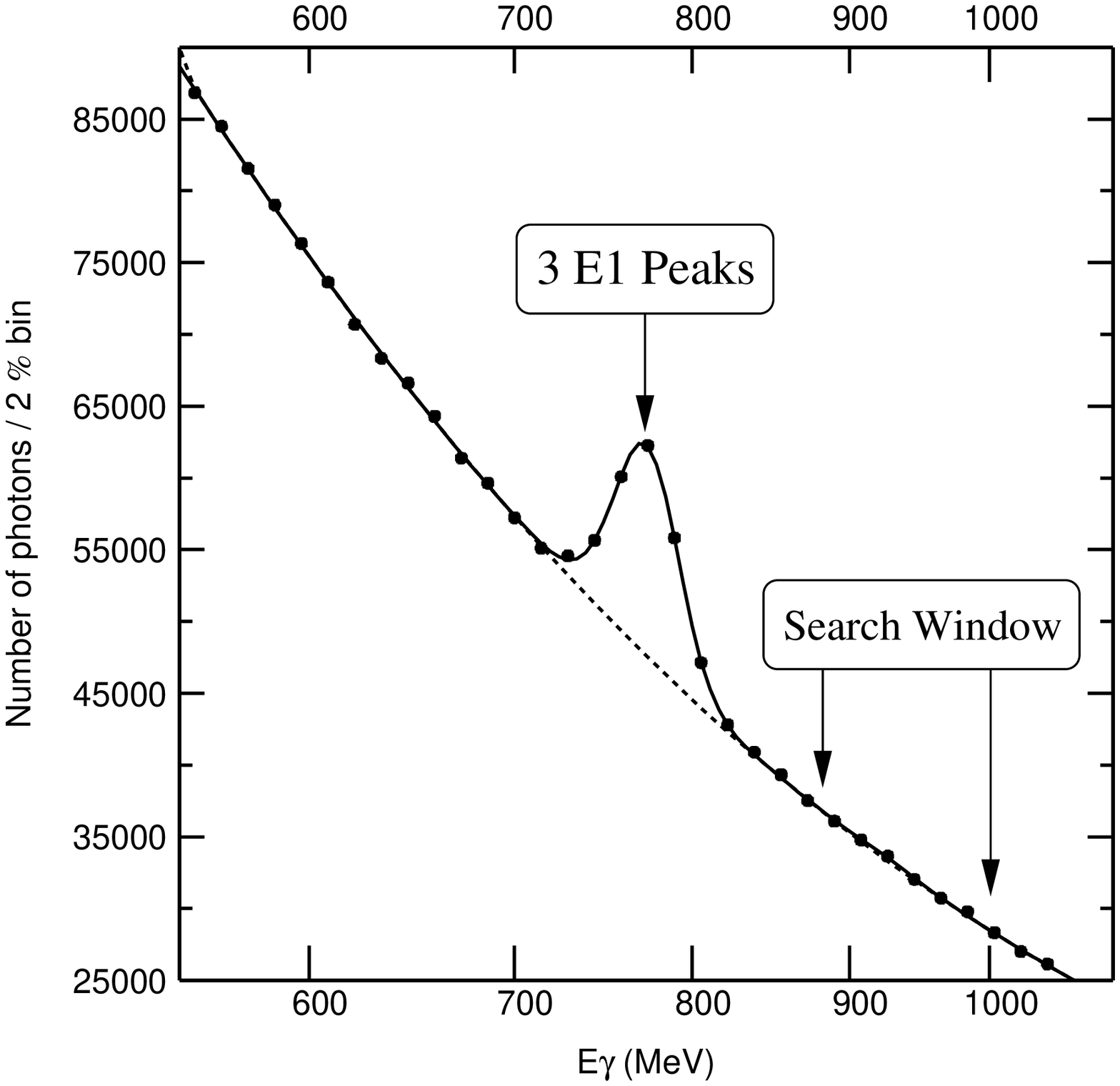,width=4.0in}}
\scalebox{0.8}{\epsfig{file=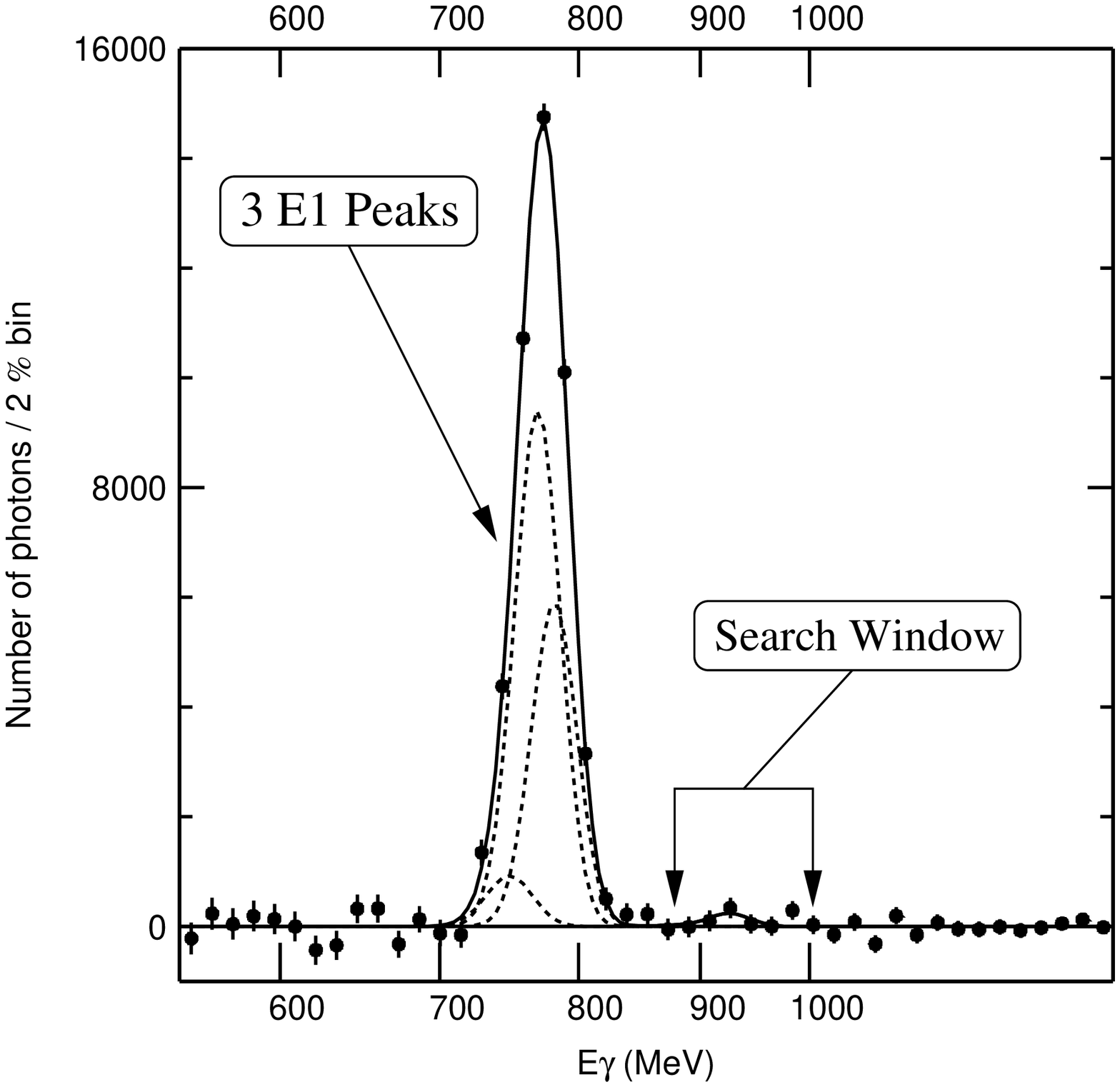,width=4.0in}}
\caption{The inclusive photon spectrum in the E1 peak ($\chi_b(2P_J)\to\gamma\Upsilon(1S)$) and the $\eta_b$ search region (left) and the background-subtracted spectrum (right).}
\label{fig:etabslope}
\end{center}
\end{figure}

\begin{figure}[ht]
\begin{center}
\scalebox{0.8}{\epsfig{file=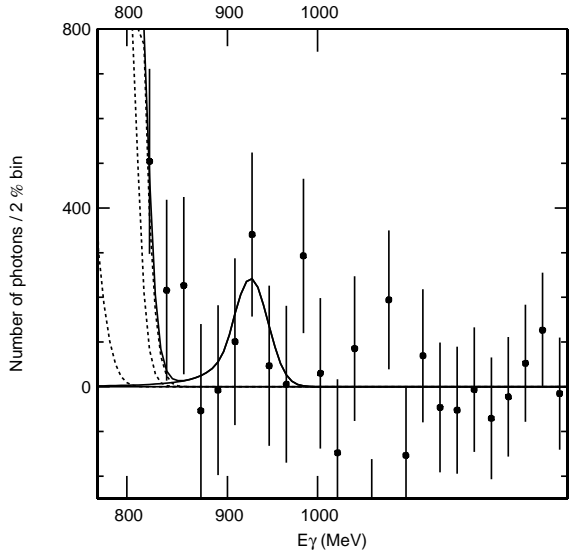,width=4.0in}}
\caption{The inclusive photon spectrum in the E1 peak ($\chi_b(2P_J)\to\gamma\Upsilon(1S)$) and the $\eta_b$ search region (background-subtracted).}
\label{fig:etab_max}
\end{center}
\end{figure}

We use logarithmic binning in energy which preserves nearly constant resolution
across a wide range of energies. Figure \ref{fig:etabslope} shows the fit. For illustrative purposes we show in Figure \ref{fig:etab_max} a close-up of the signal region (background subtracted) with this fit superimposed. We performed a series of fits assuming $E_\gamma$ = 880 to 1000 MeV, obtaining a maximum signal yield of $698\pm 463$ which is only $1.5\sigma$ significant. Since we do not find evidence for the signal, we set upper limits on ${\cal B}[\Upsilon (3S)\to \gamma \eta_{b}(1S)]$ (see Section \ref{sec:upperl}).

\section{Detection Efficiency and Branching Ratio}\label{sec:eff}

In this section we describe how the photon detection efficiency is estimated. 
We also discuss our estimates of systematic errors in the calculation of the branching ratio. 
At the end of this section, we present upper limits on the branching ratio.

\subsection{Detection Efficiency}

\begin{figure}[ht]
\begin{center}
\scalebox{.80}{\epsfig{file=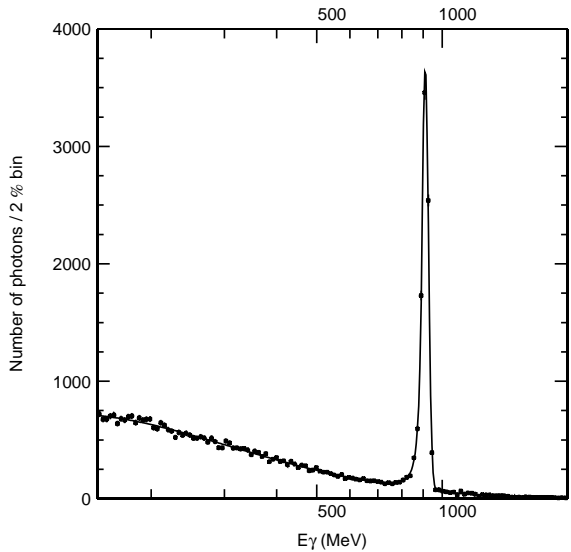,width=4.0in}}
\caption{Results of the fit to the signal Monte Carlo.}
\label{fig:etabmc}
\end{center}
\end{figure}

In order to convert the fitted amplitude to the branching ratio of $\Upsilon (3S) \to \gamma\eta_{b}(1S)$, 
we generated 18000 $\Upsilon (3S)$ decays using the LUND/JETSET with radiative tails in which $\Upsilon (3S)$ 
always decays into $\eta_{b}(1S)$ and $\gamma$ isotropically.\footnote{In our signal MC, the mass of $\eta_{b}(1S)$ was set to be 9.4GeV/$c^2$ 
and it decays into 2 gluon jets according to the LUND/JETSET model.}
The fit to the Monte Carlo photon spectrum is shown in Fig.~\ref{fig:etabmc}.
From the fit, the signal efficiency is $\epsilon_{flat}=(50.8\pm 0.7)\%$.
This efficiency is further corrected to account for the true angular distribution of M1 photons, $1+\cos^2{\theta}$ \cite{xbal}. This function is integrated over the barrel region of the calorimeter, a corrected efficiency gives $\epsilon=0.91\cdot \epsilon_{flat}$.

\subsection{Systematic Study}

To allow for the possible mismodeling of the efficiency of various cuts, we calculate ${\cal B}(\chi_{b}(2P_{J}) \to\gamma \Upsilon (1S))$ using the same method as above, and then using the same method without the $\pi^0$ suppression and $\gamma$ isolation cuts. Variations in the branching fractions calculated in these ways is taken as the systematic error.

In order to estimate the systematic error on the signal efficiency due to the Monte Carlo modeling of the decay, we generate signal Monte Carlo with a different $\eta_b$ decay model \cite{qq}.
We then compare the efficiencies derived from these two models.

The number of $\Upsilon (3S)$ decays in our sample is determined with 2\% systematic error due to the uncertanities in the hadronic selection efficiency

We summarize our study of systematic error in Table \ref{Ta:syst}.
The total systematic error is obtained by summing all entries in quadrature.

\begin{table}[ht]
\caption{Systematic Errors in ${\cal B}$}\label{Ta:syst}
\begin{center}
\begin{tabular}{lc}
Description & Error \\
\hline
Photon Isolation Cuts&  1.9\% \\
$\pi^{0}$ Suppression Cuts& 1.1\% \\
Signal Efficiency& 4.8\% \\
Hadronic Event Selection Efficiency& 2.0\% \\
\hline
TOTAL                      & 5.6\% \\
\end{tabular}
\end{center}
\end{table}

\subsection{Upper Limit on Branching Ratio}\label{sec:upperl}

We obtain a 90\% confidence level upper limit on the fitted amplitude of signal candidates by the following method. We only consider the \emph{physical region} of the signal amplitude assuming that the shape of the likelihood is Gaussian. We then determine the 90\% confidence level upper limit to be the point below which 90\% of the \emph{positive side} of the likelihood distribution is included ({\it e.g.} see \cite{pdgul}). Once we find the upper limit of the amplitude, we then convert it to a branching ratio by dividing by the signal efficiency and the number of $\Upsilon (3S)$.
Finally we accommodate our systematic errors by ${\cal B}_{UL}={\cal B}_{original}\times 1.056$.

\begin{figure}[ht]
\begin{center}
\scalebox{0.95}{\epsfig{file=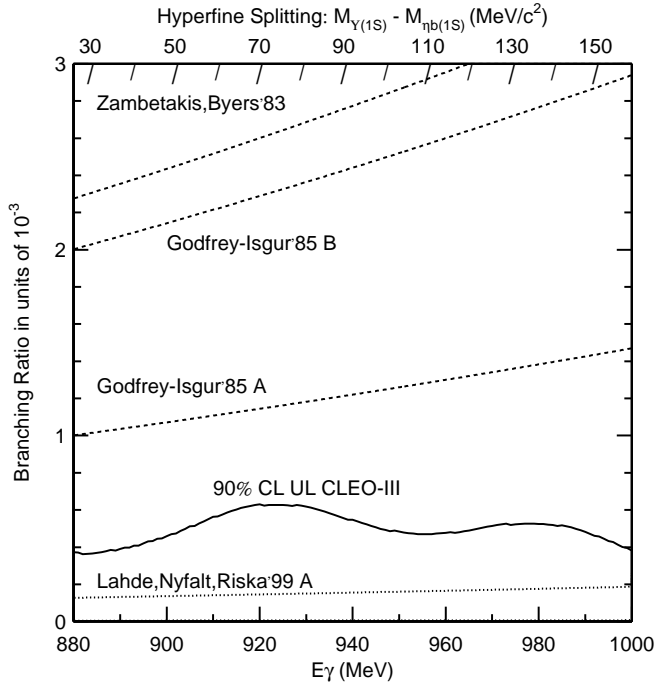,width=4.0in}}
\caption{Upper Limit on ${\cal B}$[$\Upsilon (3S)\to \gamma \eta_{b}(1S)$]. The model predictions are taken from Ref. [1].
They are extrapolated to various singlet-triplet splitting values
 (corresponding to various signal photon energies) via the phase space factor
($E_{\gamma}^3=k^3$).}
\label{fig:ul}
\end{center}
\end{figure}

Figures \ref{fig:ul} shows our upper limits on ${\cal B}[\Upsilon (3S)\to \gamma \eta_{b}(1S)]$, as a function of the emitted photon energy, $E_\gamma$ (equivalently the hyperfine splitting = $M_{\Upsilon (1S)}-M_{\eta_b(1S)}$), in which we searched for the signal between 880MeV and 1000MeV by 
repeating the above calculations of upper limit branching ratios for each fixed signal photon energy. Some model predictions taken from Ref. \cite{Godfrey} are also overlaid in the plot.

\section{Conclusions}\label{sec:concl}

Using $4.7$ million $\Upsilon (3S)$ decays 
collected by the CLEO III detector, we have searched for the ground state of the $b\bar{b}$ system, $\eta_{b}(1S)$, through the hindered Magnetic Dipole (M1) transition, $\Upsilon (3S)\to \gamma \eta_{b}(1S)$. 
No significant signal was found in the range of 880MeV$\leq E_{\gamma} \leq$ 1000MeV. 
We set 90\% confidence level upper limits on ${\cal B}[\Upsilon (3S)\to \gamma \eta_{b}(1S)]$  in this photon energy range
which rule out some previously published phenomenological predictions of this rate.
\\[15pt]
\begin{center}
\begin{tabular}{c}
\large Acknowledgments \\
\end{tabular}
\end{center}
\noindent We gratefully acknowledge the effort of the CESR staff in providing us with excellent luminosity and running conditions.
M. Selen thanks the PFF program of the NSF and the Research Corporation, 
and A.H. Mahmood thanks the Texas Advanced Research Program.
This work was supported by the National Science Foundation and the
U.S. Department of Energy.
\nopagebreak

\end{document}